\documentclass[11pt]{article}
\def\bec{\begin{center}}
\def\enc{\end{center}}

\def\ben{\begin{equation}}
\def\ba{\begin{array}}
\def\bea{\begin{eqnarray}}

\def\een{\end{equation}}
\def\eea{\end{eqnarray}}
\def\ea{\end{array}}
\def\btab{\begin{table}}
\def\btabu{\begin{tabular}}
\def\etab{\end{table}}
\def\etabu{\end{tabular}}
\def\bit{\begin{itemize}}
\def\eit{\end{itemize}}
\def\bef{\begin{figure}[htb]}
\def\befh{\begin{figure}[!h!]}
\def\enf{\end{figure}}

\def\la{\langle}
\def\ra{\rangle}

\def\a{\alpha}

\def\bsig{\mbox{\boldmath $\sigma$}}
\def\gb{\beta}

\def\D{\Delta}
\def\d{\delta}

\def\L{\Lambda}

\def\l{\lambda}

\def\O{\Omega}

\def\S{\Sigma}
\def\s{\sigma}

\def\Ga{\Gamma}

\def\b1{{\bf 1}}

\def\bme{\mbox{\boldmath $e$}}

\def\br{\mbox{\boldmath $r$}}
\def\brp{\mbox{\boldmath $r$}^\prime}

\def\bJ{\mbox{\boldmath $J$}}

\def\bL{\mbox{\boldmath $L$}}

\def\bA{\mbox{\boldmath $A$}}
\def\bB{\mbox{\boldmath $B$}}

\def\barQ{\bar{Q}}

\def\nn{\nonumber}
\def\bb{\left(}
\def\eb{\right)}
\def\bs{{\bf s}}
\def\br{{\bf r}}
\newcommand{\name}{\arabic{section}}
\newcommand{\newsection}[1]{\section{#1}\renewcommand{\theequation}
                              {\name.\arabic{equation}}
                            \setcounter{equation}{0}}

\def\bs{{\bf s}}

\def\rw{\rule[-5mm]{0mm}{12mm}}
\def\rw0{\rule[0mm]{0mm}{15mm}}
\def\rb{\raisebox{3mm}[0pt]}
\def\rb0{\raisebox{0mm}[0mm][20truemm]{\null}}
\usepackage{epsfig,latexsym}

\begin{document}
\title{Spin Effects in Heavy Hybrid Mesons on an Anisotropic Lattice}
\author{{\bf  I.T. Drummond, N.A. Goodman, R.R. Horgan, H.P. Shanahan, L.C. Storoni }\\
        \\
        D.A.M.T.P.\\ 
        Silver Street, Cambridge, England CB3 9EW\\
        }
\maketitle
\begin{abstract}
We present a quenched calculation of lowest bottomonium hybrid states
on an anisotropic lattice with Landau mean-link tadpole improvement, using the improved 
NRQCD Hamiltonian. We investigate the quark-spin $s=0,1$ sectors 
which contain both the non-exotic $1^{--}$ and the exotic $1^{-+}$, and demonstrate 
their degeneracy in the case of the lowest order Hamiltonian. Both states are 
found at around 1.6 GeV above the $\Upsilon$ ground state. We examine the 
spin-splitting for several hybrid states ($1^{--}$, $1^{-+}$, $0^{-+}$, $2^{-+}$)
which is due to the $-c_1\,\bsig . \bB/2m_b$ term in the NRQCD
Hamiltonian, all other terms having negligible effect. The spin splittings 
are well resolved outside errors and are surprisingly large. We investigate their
dependence on $c_1$ for several values of $c_1$. We calculate one contribution to
these splittings using the Born-Oppenheimer picture for hybrids and show that
the observed size of the effect is plausibly explained by mixing with hybrid states 
with more than one constituent gluon.

\end{abstract}
\vskip 10 true mm
\begin{center}
keywords: lattice, gauge, anisotropic, QCD
\end{center}
\vfill
DAMTP-1999-178\\
\newpage
\newsection{\label{introduction}\bf Introduction}

The spectrum of hybrid mesons provides fresh insight into the structure
of QCD. Hybrid mesons are therefore of intense interest, both
experimentally and theoretically. In this paper we study {\it heavy} hybrids
using lattice QCD. The results invite testing in future $B$-factory
experiments.

Because of the large mass of the $b$-quark we can apply the NRQCD
approach, which has been successful for the $\Upsilon$-system \cite{davies,drummond,aliea},
with some confidence. Indeed we expect the quarks in heavy hybrid states 
to be even more slowly moving than is the case for the $\Upsilon$ itself.
We work in the quenched approximation for the gauge fields. However it 
is unlikely that the use of dynamical gauge configurations would radically 
change our results.

In order to reconcile computational efficiency with the need for accurate 
measurement of the relatively large values of hybrid excitations, we use 
an anisotropic lattice that is spatially coarse but refined in the time direction. 
The gauge field configurations are determined by the spatially improved
gluon field action \cite{alea},
\ben
S_n=-\beta\sum_{x,s>s'}{\chi^{-1}\left\{\frac{5}{3}\frac{P_{s,s'}}{u_s^4}-\frac{1}{12}\frac{R_{ss,s'}}{u_s^6}-\frac{1}{12}\frac{R_{s's',s}}{u_s^6}\right\}}-\beta\sum_{x,s}{\chi\left\{\frac{4}{3}\frac{P_{s,t}}{u_s^2u_t^2}-\frac{1}{12}\frac{R_{ss,t}}{u_s^4u_t^2}\right\}}
\een
where $s,s'$ run over spatial links, $P_{s,s'}$ and $P_{s,t}$ are $1\times 1$ plaquettes, 
$R_{ss,s'}$  and $R_{ss,t}$ are  $2\times 1$ plaquettes, $\chi$ is the bare anisotropy. 

The tadpole parameters $u_s$ and $u_t$ are defined to be respectively, the expectation 
values of the traced spatial and time-like link matrices in the Landau gauge \cite{lepa0}.
The actual values are established self-consistently by an appropriate
iteration procedure. The choice of Landau mean-link tadpoles is motivated by evidence that 
their use leads to  better continuum behaviour than can be obtained from 
plaquette tadpoles \cite{shtr0,shtr1,alea0}.

The renormalized anisotropy is $\chi_R = a_s/a_t$ where $a_s$ and $a_t$ are the
spatial and time-like lattice spacings respectively. The value of $\chi_R$ can
be obtained in several different ways. In a separate paper we have determined
it by fitting torelon and glueball dispersion relations measured on the lattice 
and by measuring the heavy quark potential in both the fine and coarse directions \cite{alea0}.
In this paper we use the bare parameters $\gb=1.8$ and $\chi=6$ together with the
appropriately tuned tadpole parameters $u_s=0.7216$ and $u_t=0.99208$, with corresponding
value $\chi_R=5.32(2)$~. From the $\Upsilon(1P-1S)$ mass gap measured on the lattice
we find $a_t^{-1}=4503(46)$ MeV and $a_s^{-1}=819(9)$ MeV.

Our calculations are a continuation of previous work \cite{maea0,maea1} in which we investigated 
the effect of including all terms in the quark NRQCD Hamiltonian up to $O(mv^6)$~. 
However in measuring the heavy hybrids we have found that terms of $O(mv^4)$ and 
higher have negligible effect on our results. We therefore drop such terms in order 
to reduce the computational load. This circumstance is consistent with the observation 
that the heavy quarks move rather slowly in the hybrid state. 

The contribution from the first spin dependent term is large however and central 
to our results. Accordingly we retain this term and the $O(a_s)$ improvement to
$D^2$, and use the quark 
Hamiltonian $H = H_0+\d H$, where
\bea
H_0&=&-\frac{D^2}{2m_b} \nn\\
\d H&=&-c_0\frac{\Delta^4}{8m_{b}^3}-c_1\frac{g}{2m_b}\bsig.{\bB}
    + O(mv^4)~.\label{h_nrqcd}
\eea
The $\bB$-field is constructed from an improved form for $F_{\mu\nu}$ \cite{alea2}.
We take the tree-level value $c_0=1$~. 

The thrust of our investigation is to study the dependence of the hybrid 
mass spectrum on $c_1$. This parameter may be subject to a large renormalization
that forces its departure from the tree level value $c_1=1$~.  
The states considered are the low-lying magnetic hybrids which
are classified as the $1^{--}$, which has quark spin $s=0$, and the 
$0^{-+},~1^{-+},~2^{-+}~$, which have $s=1$. The $1^{--}$ is non-exotic 
and is the only one of these states that can be produced directly in $e^+e^-$ 
collisions. Of the others, the $1^{-+}$ is exotic. 

In section \ref{method} we outline the NRQCD approach to calculating heavy 
hybrid states, and discuss the operators and smearing used. In section 
\ref{results} we discuss finite-size effects and present the results from
our simulation. In section \ref{BOP} we discuss the bag model calculation of spin
effects and inter-state mixing in the Born-Oppenheimer approximation. In section
\ref{discussion} we present our discussion and draw our conclusions.

\newsection{\label{method}\bf The NRQCD approach}

The NRQCD approach to computing the heavy hybrid propagator is well known 
The Euclidean time evolution equation for the quark
propagator $G({\bf x},t;{\bf y})$ is computed according to the prescription:
$$
G({\bf x},t+1;{\bf y}) =
\left(1-\frac{a_tH_0}{2n}\right)^nU^{\dagger}_t({\bf
x})\left(1-\frac{a_tH_0}{2n}\right)^n(1-a_t\delta H)G({\bf x},t;{\bf y})\:\:\:t>0~,
$$
where $H_0$ and $\d H$ are given in eq(\ref{h_nrqcd}) and $n=2$ for our simulations. 
The quark propagator
is then determined by its initial distribution, $G({\bf x},t=0;{\bf y}) = \Ga(\bf x,\bf y)$~. 
The choice of initial distribution is governed by the use to be made of the propagator. 
We use propagators computed from the following in initial distributions:
\bea
\Ga^L({\bf x,y})&=&\delta^{(3)}(\bf x,y)~,\nn\\
\Ga^S({\bf x,y})&=&\left(1+\frac{l^2D^2}{4m}\right)^m \delta^{(3)}(\bf x,y)~,\nn\\
\Ga^P_i({\bf x,y})&=&\left(1+\frac{l^2D^2}{4m}\right)^m D_i\delta^{(3)}(\bf x,y)~,\nn\\
\Ga^H_i({\bf x,y})&=&D^2\left(1+\frac{l^2D^2}{4m}\right)^m B_i\delta^{(3)}(\bf x,y)~,
\label{smearing}
\eea
where $D_i$ is a (discrete) covariant derivative. The propagator resulting from the
local source, $\Ga^L$, is combined with the each of the other smeared propagators, 
after hermitian conjugation, and traced with appropriate spin matrices to create the
correlators for the set of operators listed in Table \ref{states}. 

The correlators
appropriate to hybrid states are computed from $\Ga^H_i$~. An analysis of hybrid states 
in the Born-Oppenheimer approximation indicates that the quark and anti-quark
tend to remain apart from one another because  they experience an angular momentum barrier 
and a color repulsion (the $q\bar{q}$ system is in a color octet). 
The ``donut'' smearing incorporated in $\Ga^H$ through the application of a final $D^2$, 
was successfully invoked in order to improve the overlap of the operators with the widely
separated $q\bar{q}$ system in the hybrid.  

The forms of the operators we use to interpolate the magnetic hybrid states
are shown in Table \ref{states}.

\btab[htb]
\bec
\btabu{|c|c|c|c|c|c|}\hline
&State&s&l&j&$J^{PC}$\\\hline
spin-0 S-wave&$\bar{q}q$&0&0&0&$0^{-+}$\\\hline
spin-1 S-wave&$\bar{q}\s_iq$&1&0&0&$1^{--}$\\\hline
spin-0 P-wave&$\bar{q}D_iq$&0&1&0&$1^{+-}$\\\hline
&$\bar{q}\s_iD_iq$&1&1&0&$0^{++}$\\
spin-1 P-wave&$\bar{q}\epsilon_{ijk}\s_jD_kq$&1&1&0&$1^{++}$\\
&$\bar{q}(\s_iD_j+\s_jD_i-\frac{2}{3}\delta_{ij}\s_kD_k)q$&1&1&0&$2^{++}$\\\hline
spin-0 Hybrid&$\bar{q}B_iq$&0&0&1&$1^{--}$\\\hline
&$\bar{q}\s_iB_iq$&1&0&1&$0^{-+}$\\
spin-1 Hybrid&$\bar{q}\epsilon_{ijk}\s_jB_kq$&1&0&1&$1^{-+}$\\
&$\bar{q}(\s_iB_j+\s_jB_i-\frac{2}{3}\delta_{ij}\s_kB_k)q$&1&0&1&$2^{-+}$\\\hline
\etabu
\enc
\caption[]{\label{states}\small
The forms of the operators used to interpolate the magnetic hybrid states we
study and their $J^{PC}$ classification. Here  $l$ and $s$ are the $Q\barQ$
angular momentum and spin, and $j$ is the gluon angular momentum.
}
\etab

The charge conjugation $(C)$ and parity $(P)$ of the quarkonia and hybrids, satisfy 

\bec
\btabu{cccc}
&pure&magnetic&electric\\
&quarkonium&hybrids&hybrids\\
&\parbox{30mm}{~}&\parbox{30mm}{~}&\parbox{30mm}{~}\\
$C=$~~~&$(-1)^{l+s}$&$(-1)^{l+s+1}$&$(-1)^{l+s+1}$\\
\\
$P=$~~~&$(-1)^{l+1}$&$(-1)^{l+j}$&$(-1)^{l+j+1}$\\
\etabu
\enc
where $l$ and $s$ are the $\bar q q$ angular momentum and spin and $j$
is the gluon angular momentum. The S,P and some of the hybrids are
listed in Table \ref{states}. 

\newsection{\label{comp}\bf Computational details}

In the source and sink operators (Table \ref{states}) all links were smeared using
covariant smoothing in a manner similar to that described in eq(\ref{smearing})
with an appropriately defined $D^2$. The $\bB$ fields were computed from the smeared 
links and were themselves subsequently smeared in a similar manner.  
The range for the smearings was typically $l=2$ on gauge links and $l=3$ on 
the quark source. The effectiveness of the smearing was confirmed by a reduction
in statistical errors by a factor of three or more. To reduce statistical errors 
to within 1\% we generally required between 40,000 and 60,000 independent correlators.
We used a 50 sweep Cabbibo-Marinari thermalisation to
generate the initial gauge fields, and then employed 10 sweeps between
measurements. 

The computations were performed
on the Hitachi SR2201 of the Cambridge High Performance Computing Facility
and at University of Tokyo Computer Centre. They were farmed across 64 processors
each of which returned statistically independent results that were used to
construct the full covariance matrix. In the standard way we 
performed a correlated fit to the correlation functions of the relevant
operators in order to determine the masses of the quarkonium and hybrid excitations.
We confirmed that 10 sweeps between measurements was sufficient for decorrelation.

The timings for the calculation was such that for the hybrids an 8 hour 
job produces about 80 correlators per processor for
the $6^3\times 36$ lattice, about 25 for the $8^3\times 48$ and 5
for the $10^3\times 60$. 

\newsection{\label{finite}\bf Finite size effects}

We first examined the sensitivity of the calculation to finite size effects. 
In Table \ref{finite_size} we show the mass gap from the $\Upsilon(1S)$ to the 
$1^{--}$ hybrid (without spin effects) measured by us on lattices $L^3\times T$ 
for a range of values of $L$~. Only when $L$ is reduced down to 4 is there any 
detectable effect from the finite spatial lattice volume.  This is consistent 
with the results of Manke et al \cite{maea0,maea1}~. 

Since a hybrid is typically $2-3~\mbox{GeV}^{-1}$ \cite{haea} across, its size is 
$\sim~4a_s$. For $L=6$ it covers a substantial part of the lattice volume. It is 
surprising therefore that finite size effects are so small since we might expect that
a quark would react more strongly to the nearby image of an anti-quark across the 
boundary of the spatial lattice than to the more distant anti-quark within
the the spatial volume. However a strong effect depends on a transition in which
the gauge field flux tube switches from the original quark anti-quark linkage
to one crossing the spatial boundary. The corresponding transition amplitude
involves the expectation value of a closed Polyakov loop that encircles the torus. 
This vanishes in the confining phase because of the $Z_3$ centre symmetry 
of the action.  We argue therefore that our results are consistent
with a picture in which the major interaction across the boundary of the spatial box is a 
weak Van der Waals-type force. 

There are other re-linking transitions, in the case of $SU(3)$, involving the multiple wrapping
of flux lines round the torus, that do not vanish. However we expect them to be suppressed
because of the high energy of the associated flux tubes.

For the rest of our investigation we worked with $L=6$ as the most economical
choice of lattice consistent with negligible finite size effects.

\newsection{\label{results}\bf Results}

We use as an energy baseline the {\it spin averaged} $\Upsilon(1S)$ state.
Table \ref{upsilon}  shows our results for the dependence of this baseline on $c_1$~. 
They are consistent, with quadratic behaviour in $c_1$ reflecting the fact that the shift 
is due to a quark mass renormalization.

Since the (spin averaged) $1S-1P$ mass difference $\D M_{PS}$ in lattice units is known 
to be insensitive to the quark mass $m_b$ we use it to determine the temporal lattice spacing 
in physical units. From the assignment $\D M_{PS}=440$ MeV we infer that 
$a_t^{-1}=4503(46)~\mbox{MeV}$~. From other work \cite{alea0} we know that $\chi_R=5.32(2)$ 
yielding $a_s^{-1}=819(9)~\mbox{MeV}$~. We determine $m_b$ by measuring the mass of the 
$\Upsilon(1^3S)$ state from its dispersion relation and then requiring it to be consistent 
with the mass measured directly. We find $m_b=4.75~\mbox{GeV}$~.

\btab
\bec
\btabu{|c|c|c|c|c|}\hline
~(L,T)~&(4,50)&(6,36)&(8,48)&(10,60)\\\hline
$(M_H-M_{\Upsilon})a_t$&0.281(7)&0.244(7)&0.244(9)&0.247(9)\\\hline
\etabu
\enc
\caption[]{\label{finite_size}\small 
Dependence of the mass, $M_H$, of the $1^{--}$ hybrid state on the finite
size, $L$. In all cases $c_1=0$, $a_t^{-1} = 4503(46)~\mbox{MeV}$.
}
\etab

Table \ref{hybrids} shows our results for the masses of the hybrid states. 
We also show here the behaviour of the mass with
variation of the $c_1$ co-efficient in $\delta H$, which controls the
magnitude of the $\bsig.{\bf B}$ term. These results are illustrated in Fig. \ref{masses}.

\btab
\bec
\btabu{|c|c|c|c|}\hline
$c_1$&$1S_3^1$&$1S_1^0$&$\la 1S \ra_{spin}$\\\hline
0&0.1237(7)&0.1246(6)&0.1239(5)\\\hline
1&0.1196(4)&0.1143(5)&0.1183(3)\\\hline
2&0.1054(7)&0.086(3)&0.1005(3)\\\hline
\etabu
\enc
\caption[]{\label{upsilon}\small 
Mass of the $1 S_3^1,~1S_1^0$ and the spin-averaged value for different values
of $c_1$.
}
\etab

\btab[htb]
\bec
\btabu{|c|c|c|c|c|c|}\hline
State&$c_1$&$\chi^2/{d.o.f}$&$a_tE_H$&$a_t(E_H-E_{Q\barQ})$&$E_H-E_{Q\barQ}$
MeV\\\hline
$0^{-+}$&1&0.91&0.444(9)&0.331(9)&1490(43)\\
&2&0.34&0.32(1)&0.22(1)&991(46)\\\hline
$1^{--}$&0&0.98&0.491(5)&0.367(5)&1653(27)\\
&1&0.48&0.4815(43)&0.3632(43)&1635(25)\\
&2&0.85&0.4702(65)&0.3648(65)&1643(33)\\\hline
&-2&0.9&0.486(6)&0.385(6)&1734(32)\\
&-1&1.02&0.505(5)&0.387(5)&1743(28)\\
$1^{-+}$&0&0.26&0.4887(50)&0.365(5)&1644(27)\\
&1&0.90&0.4540(60)&0.336(6)&1513(31)\\
&2&0.5&0.382(5)&0.281(5)&1265(26)\\\hline
$2^{-+}$&1&0.63&0.516(6)&.398(6)&1792(32)\\
&2&0.96&0.488(7)&.388(7)&1747(36)\\\hline
\etabu
\enc
\caption[]{\label{hybrids}\small
Mass of the hybrid states with variation of the spin coupling
magnitude, accurate to $O(mv^6)$ with $\beta=1.8$, $am_b=4.75$,
$a_t^{-1}=4.503(46)$ GeV.}
\etab
 
For each choice of $c_1$ (defined in eq(\ref{h_nrqcd})) we compute the 
$s=0,1,~l=0$ quarkonium correlators and the hybrid correlators for the
states listed in Table \ref{states}. Although, in principle, we could 
consider a matrix of source-sink operator pairs and fit the corresponding
matrix correlator, we found it sufficient to consider the one
correlator for which both source and sink were smoothed as described above.
We found that the $S$-wave quarkonium correlators were well fitted
with two-exponentials over the range (in lattice units) $10 \le t \le 50$.
The hybrid states are much heavier and the signal for the propagator
was too noisy for $t > 15$ but we obtained excellent fits with a single exponential
for typically $t > 5$. An representative effective mass plot for the $1-+$ for $c_1=-1$
is shown in Fig. \ref{effm} which has a plateau extending from $t=4-10$. The dashed
lines shown the upper and lower error bounds from the fit. 

The $1^{--}$ hybrid is of particular interest since it will appear as an intermediate state
in $e^-e^+$ scattering at $B$-meson factories. It is noteworthy that its
excitation energy above the ground state is unchanged by the spin interaction
independently of its strength. The absence of a term $O(c_1)$ implies that
there is no direct energy shift induced by the spin term in the Hamiltonian 
and the absence of a term $O(c_1^2)$ implies that there is no mixing with quarkonia
even though this is possible on the basis of quantum numbers. We explain below why we 
feel that this result is compatible with the bag model of hybrid states. Our prediction 
for its mass is 1644(17) MeV above the ground state. This is compatible with, if somewhat
higher than, the value from some other studies \cite{maea0,maea1,maea2,juea}. It would be of great 
interest to observe experimentally a state at this energy level.

In contrast the masses of the spin one states are strongly affected by the spin term in the
NRQCD Hamiltonian. Of these the $1^{-+}$ is exotic and cannot mix with quarkonium states.
For this state we have investigated the dependence of its mass on $c_1$ for both positive
(physical) and negative (unphysical) values. This allows us to disentangle contributions
that are linear and quadratic in $c_1$. From Table \ref{hybrids} we see that the effect
of the spin term is large. For the tree level value $c_1=1$ the $1^{-+}$ is depressed
by $\sim 130$ MeV. A large renormalization of $c_1$ because of radiative corrections
means that $c_1=2$ is a possibility, a value which causes the $1_{-+}$ to be depressed by
almost $300$ MeV to 1265(26) MeV above the spin-averaged $1S$ $Q\barQ$ state. A simple 
fit to the $1^{-+}$
masses is 
\ben
a_t(E_H-E_{Q\barQ})~=~0.369(4) - 0.026(2)c_1 - 0.09(1)c_1^2~,\label{fit}
\een
with $\chi^2=0.3$. 
These effects are unexpectedly large since spin effects depend on the quark velocity which 
is low in hybrid states.

The non-exotic $0^{-+}$ and $2^{-+}$ states can mix with quarkonium states and, in principle, 
there can be contamination of the signal which would make these states hard to pick out in
the simulation. However, as in the case of the $1^{--}$ we find no such problem. On the basis
of a bag model calculation this mixing is indeed found to be very small and certainly at a 
level of less than 1\% in our simulations. The $0^{-+}$ is depressed even more than the 
$1^{-+}$ and for $c_1=2$ is only $\sim 1000$ MeV above the $Q\barQ$ ground state. The $2^{-+}$ 
rises for $c_1 > 0$ although the magnitude of the shift is not as great as that for the
other states. This is compatible with a change of sign of the coefficient of $c_1$ in
the fit eq(\ref{fit}) which causes this term to compete with the quadratic contribution.
We shall argue in the next section that at least a part of the coefficient of the term 
linear in $c_1$ is proportional to $\bL\cdot\bs$ which does indeed have this qualitative 
behaviour. We have not investigated the $0^{-+}, 2^{-+}$ for $c_1 \le 0$.

\newsection{\label{BOP} Born-Oppenheimer Model}

Hasenfratz et. al \cite{haea} and Kuti and Morningstar \cite{juea} have stressed the usefulness of 
a Born-Oppenheimer picture of hybrid states in the bag model of hadrons. We suggest that 
our results can be consistently interpreted in this approach. We investigate the effect
on mass shifts of a gluon exchanged between the quark and the anti-quark, and the 
mixing effect induced by transitions to a two-gluon state. We find the former
to be very small while the latter can provide a sizeable contribution that goes
some way to explaining our observations on the mass shifts of the hybrid states.
We also estimate the mixing of the hybrid and quarkonia states and find this to
be negligibly small for all reasonable values of $c_1$.

For simplicity we assume that the bag is spherical. The wavefunction for a hybrid of 
total orbital angular momentum $(L,m)$ in the 
BO picture is
\ben
\Psi^\L_{Lm}(\br,\brp_g) = f^L_{j\L}(r)\,{D^{*L}_{m\L}}(\O)\bA^\l_{j\L}(\brp_g)~, \label{bostate}
\een
where $\br$ is the quark-antiquark separation vector with orientation $\O=(\theta,\phi)$,
$\brp_g$ is the gluon position vector in the body frame, $j$ is the angular momentum of 
the gluon mode with projection $\L$ onto the body-fixed z-axis, and $\l=0,1$ label 
TM and TE modes, respectively. The total angular momentum is $\bJ=\bL + \bs$. States of
definite parity are given by
\ben
\Psi^{|\L|\eta}_{Lm}~=~{1\over \sqrt{2}}\bb \Psi^{|\L|}_{Lm} + 
                                            \eta \Psi^{-|\L|}_{Lm} \eb~.\label{boparity}
\een
The parity ($P$) and charge-conjugation ($C$) are given by
\ben
P~=~\eta(-1)^{L+\l+1}~,~~~~~~~~~~~C~=~\eta(-1)^{L+j+s+1}~. \label{PC}
\een

Note, that these assignments do not agree with those quoted in the original reference
\cite{haea}, but they do agree with Le Yaouanc et. al \cite{yaea} and Rakow \cite{rapc} who remarked 
on the discrepancy.  We have carefully checked that these formulae are correct. The important 
point is that the $PC$ classification is independent of $\L$. 
States with $|\L| = 0,1,2,\ldots$ are labelled respectively by $\S,\Pi,\D,\ldots$~. The 
lowest lying states include those with $L=1,j=1$ on which we shall concentrate.   
There are both $\S$ and $\Pi$ states with the same $J^{PC}$ quantum numbers, namely,
$1^{--}~$ with $s=0$ and $0^{-+},1^{-+},2^{-+}~$ with $s=1$.

There are two simple processes resulting from the $\bsig .\bB$ interaction, 
gluon exchange between the quarks, and gluon exchange between the hybrid gluon
and one or other of the quarks. The former is a weak effect because it involves the 
behaviour of the $Q\barQ$ wavefunction near the origin where it is suppressed both by 
the angular momentum barrier and a repulsive potential due to the octet nature of the
$Q\barQ$ state. An estimate yields a result of order 1 MeV or less.

The large effect is due to gluon exchange between the constituent gluon and the quarks.
This effect is large for two reasons: the colour factors are large and it depends on the
quark wavefunction in the bulk and not on its value at the origin. We outline the
calculation below and follow the formalism of Close and Horgan \cite{clho0,clho1} and Barnes, Close
and Monaghan \cite{baea0,baea1}. The latter authors present a calculation of this spin effect in
the MIT bag model and we do not reproduce all the details here. The calculation is 
for the contribution of second-order perturbation theory to the mass shift. The intermediate
states are summed over and, in our calculation, are BO states with two constituent gluons. 
We examine the contribution from one intermediate state with $j_1=j_2=1,~\l_1=\l_2=1$. 
the diagram for this process is shown in Fig. \ref{tgluon} and corresponds to the 
exchange of a $j=1$ electric gluon. In this case, it turns out that $L^\prime = L,~s^\prime=s$ 
and $\L_1+\L_2=\L$, where $\L$ labels the initial state gluon and $\L_1,\L_2$ label the 
intermediate state gluons as shown in Fig. \ref{tgluon}. Since $\D M_H$ also does 
not depend on the sign of $\L$ we consider $\L=0,1$ which correspond to the $\S$ and 
$\Pi$ states, respectively. The bag is taken to be spherical of radius $R_B$. 
The interaction Hamiltonian and the expressions for the gluon modes, 
$\bA^\l_{j\L}$, and associated $\bB$ fields are given in \cite{baea0}. The field operators 
contributing to the mass shift are expanded on this mode basis using the usual creation and 
annihilation operators. The two-gluon intermediate state can be written in the BO approximation 
in a form similar to that in eqn. (\ref{bostate}). After some manipulations of the $D$-matrices 
involved and some algebra the mass shift takes the form: 
\ben
\D M_H~=~{iCc_1\a_s \over m_bR_B^2}\,(-1)^\L\,
\bme_\L\cdot(\bme^*_{\L_1}\wedge\bme^*_{\L_2})\,\bL\cdot\bs~, \label{dm_result}
\een
where the $\bme_{\L=0}=(0,0,1)$ and $\bme_{\L=\pm 1}=(-1,\mp i,0)/\sqrt{2}$ and $C$ is an overall 
constant which includes colour factors, the characteristic bag constants
and the radial overlap integrals for the quark wavefunctions. The bag radius is $R_B$~.

Our estimate of $C$ along the lines of \cite{baea1}, yields a relatively
large value $C \sim 1.13$~. The factor $\bme_\L\cdot(\bme^*_{\L_1}\wedge\bme^*_{\L_2})$ 
is typical of the three-gluon vertex and is discussed in \cite{baea0,baea1}. 
One important result is that $\D M_H = 0$ for $\L=0$, i.e., for the $\S$ state. 
The outcome can be summarized as:
\bea
\D M_H&=&K\,c_1\,\a_s \times\left\{\ba{ccc}
~~~~\S~~~~&~~~~\Pi~~~~&~~~~J~~~~\\
0&1&2\\
0&-1&1\\
0&-2&0\\
\ea\right. \nn \\
K&=&{1.13 \over m_bR_B^2}~. \label{result}
\eea
Our calculation is, by its very nature, approximate. For example, the bag is not spherical 
but elliptical, a correction included by Kuti and Morningstar \cite{juea} when calculating the 
$Q \barQ$ BO potentials. Including this correction will increase the effect.  
Also, the bag radius is not fixed but varies with the quark motion 
in the BO approximation. The value we adopt is therefore associated with 
the most likely quark separation and we take $R_B \sim 1.5-2~\mbox{GeV}^{-1}$. With
$m_b \sim 4.75~\mbox{GeV}$ we find $K \sim 60 - 110~\mbox{MeV}$. With $c_1=1$ and $\a_s=0.4$
we find the scale of the induced spin-orbit splitting is $\sim 25-40~\mbox{MeV}$ 
to be compared with $\sim 117$ MeV for the linear term in $c_1$ from eq(\ref{fit}). 
The value chosen for $\a_s$ is typical of bag calculations and is consistent with $\a_V(\mu)$
at a suitably chosen scale based on the B-quark mass \cite{bamo}. While our prediction is 
too small our calculation is clearly incomplete. There are many other intermediate states 
that can contribute with similar magnitude of shift. The contributions of diagrams involving 
the interaction of the quarks with the constituent gluon are large partly because the colour 
factors are large and partly because there are many such terms. At $O(\a_S)$ there is a similar 
diagram to the one analyzed with $j=2$ electric gluon exchange. At $O(\a_S^2)$ there are 
diagrams involving the gluon four-point coupling, such as shown in Fig. \ref{fgluon}, 
which has many orderings and gives contributions linear in $c_1$. There are many similar 
and other higher-order diagrams which contribute to the $c_1^2$ term. It cannot be excluded 
that there are also non-perturbative effects as well. The dependence of the induced mass shifts 
on the quantum numbers will be generally more complex than the $\bL\cdot\bs$ derived 
explicitly above, which nevertheless gives a qualitative picture of the pattern observed.  
We are therefore confident that our calculation sets a lower bound on the effect.

In the previous section we remarked that the $2^{-+}$ rises above the $c_1=0$ reference 
mass by an amount smaller in magnitude than the shift of either the $0^{-+}$ or $1^{-+}$
below this point. This could be due to a cancellation between the linear and quadratic
terms in eq(\ref{fit}). However, our operators do not distinguish between
the $\S$ and $\Pi$ states since they project out definite quark angular momentum
$l_{Q\barQ}=0$ which corresponds to an equally weighted mixture of $\L=1,0,-1$. 
Based on the potentials of Kuti et. al \cite{juea} we see that the $\S$ (which has
the same $PC$ assignments as the $\Pi$) is about 150 MeV above the $\Pi$. Since
the $\S$ is unaffected by the spin-orbit term it may be that we are seeing the
$2^{-+}$ $\S$ state and not the $\Pi$ state because this latter has risen to a higher
mass. We are testing this suggestion by constructing operators diagonal in $\L$
and the results will be presented in a subsequent publication.

In order to confirm our observation that mixing with quarkonium states is 
negligible we have estimated the mixing in the BO approximation in the bag model.
Since both source and sink contain a gluon field the gluon must be annihilated and
subsequently recreated. A diagram of the kind that is relevant is shown in Fig. \ref{mixing}. 
The states that mix must have opposite spin and so the mixing is by a spin-flip
operator, in this case it is proportional to $(\bsig_Q-\bsig_{\barQ})$. We calculate
this term for the $s=1$, ground-state, hybrids mixing with the $s=0$ quarkonium state. 
A term of this kind has been discussed earlier by Ono \cite{ono} and Le Yaouanc 
et. al \cite{yaea}. Our result is compatible with theirs and we find that the
mixing amplitude from this diagram is $\sim 10^{-4}$. Ono \cite{ono} gives $\sim 2\times 10^{-4}$. 
The hybrid--$Q\barQ$ mass difference in lattice units is $\sim 0.36$ and so the effect of 
mixing at $t=10$ is $\sim 10^{-4}\times e^{-3.6} \sim .4\%$. It is unlikely that 
mixing because of excited hybrids of the same quantum numbers will contribute 
appreciably and so we expect that mixing has negligible effect on the signal.   

We believe that we have demonstrated that spin effects in heavy hybrid mesons are
large and that the explanation is that the states expanded in the Fock basis  
contain more than one gluon constituent with an appreciable probability. Our
partial estimate based on one intermediate state is $\sim 20-30$\% of the observed
effect and shows that the order of magnitude can be explained. We are in no doubt
that the complete sum over intermediate states will contribute to the shortfall
and, because the effect at lowest non-zero order is so large, it is sensible to
suppose that higher order corrections are appreciable. The most easily observed
$1^{--}$ state is unaffected since it has $s=0$ but indirect methods of production
should eventually detect the $s=1$ states and then our observations can be tested. 

\newsection{\label{discussion}\bf Discussion and Conclusions}

Ours is the first study on an anisotropic lattice that incorporates tuned Landau tadpoles
in the action. The choice of Landau tadpoles was motivated by the observation that they 
yield better scaling behaviour for observables than plaquette tadpoles \cite{alea0}.  

We have confirmed that the hybrids are at a levels in the range 900 - 1750 MeV above the $1S$ 
spin averaged ground state. This is consistent with our previous work and, allowing for systematic 
uncertainties in $a_t$ inevitable in a quenched calculation, is consistent with other evaluations. 

The important new result of our calculation is that the effect of the spin interaction 
on the hybrid spectrum is unexpectedly large. We measure energy level splittings  $\sim 100$ MeV 
and greater. The results are exhibited in Table \ref{hybrids} for a range of values of 
$c_1$ including, in some cases, unphysical negative values that help to establish parameters 
for the analytic form of the dependence on this coefficient. 

In the absence of the spin interaction ($c_1=0$), all the states we measure are degenerate.
Because the exotic cannot mix with quarkonia, neither do the other states in the multiplet.
When the spin interaction is switched on two effects can occur, (i) a shift of
the energy levels of the hybrids and (ii) a mixing of the non-exotics with appropriate
quarkonium states. We claim that our results are consistent with a scenario in which
there is a direct effect on most of the energy levels but little mixing with quarkonia.

At tree level with no lattice artifact corrections, $c_1=1$. In deriving the NRQCD 
approximation from an interacting gauge theory we would expect corrections to 
$c_1$ of $O(\a_s(ma_s)^2)$. These could be large enough to yield $c_1\sim 2$~.
Hence, the spin effect we have simulated here may be a reasonable prediction even 
for $c_1\sim 2$~. Because the terms in the NRQCD Hamiltonian which are $O(mv^4)$
and higher have neligible effect on the hybrid levels we can expect the effect of
radiative corrections to be limited to a renormalization of $c_1$ which gives hope
that the spin effects observed here are quantitatively correct. 

It would be of great interest if any of these hybrid states could be observed although,
except in the case of the $1^{--}$ which can be directly produced in $e^+e^-$ collisions,
this would be experimentally difficult.

Future work will concentrate on using source and sink operators which have the quantum numbers
of the Born-Oppenheimer approximation. This will allow us in particular to disentangle the
$\Pi$ and $\S$ states.

\bibliography{refs}
\bibliographystyle{unsrt}

\newpage
\befh
\bec
\epsfig{file=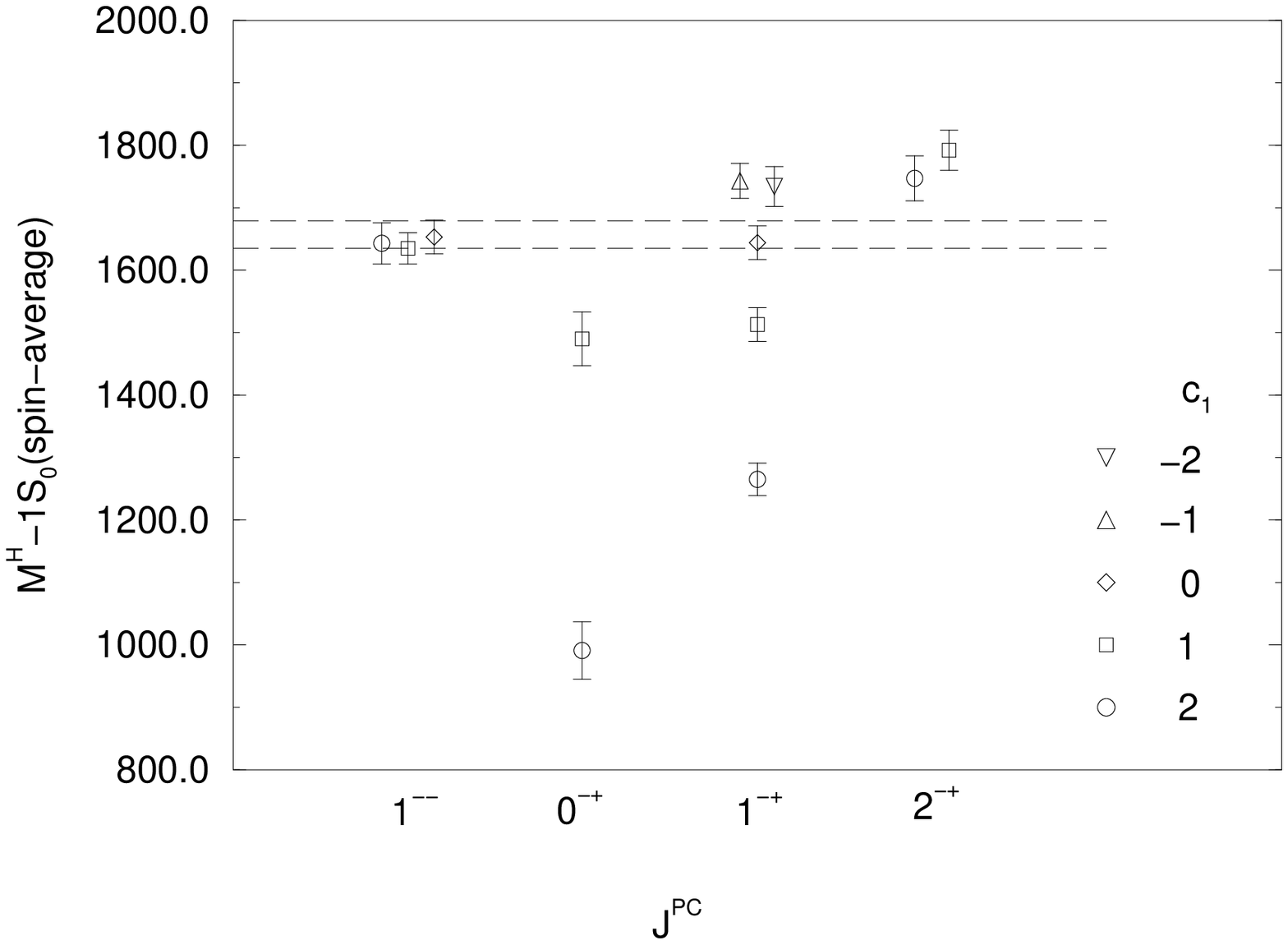,height=80mm}
\enc
\caption{\label{masses}\small
The variation of the hybrid masses with $c_1$ for the $1^{--}, 0^{-+}, 1^{-+}, 2^{-+}$.
}
\enf

\befh
\bec
\epsfig{file=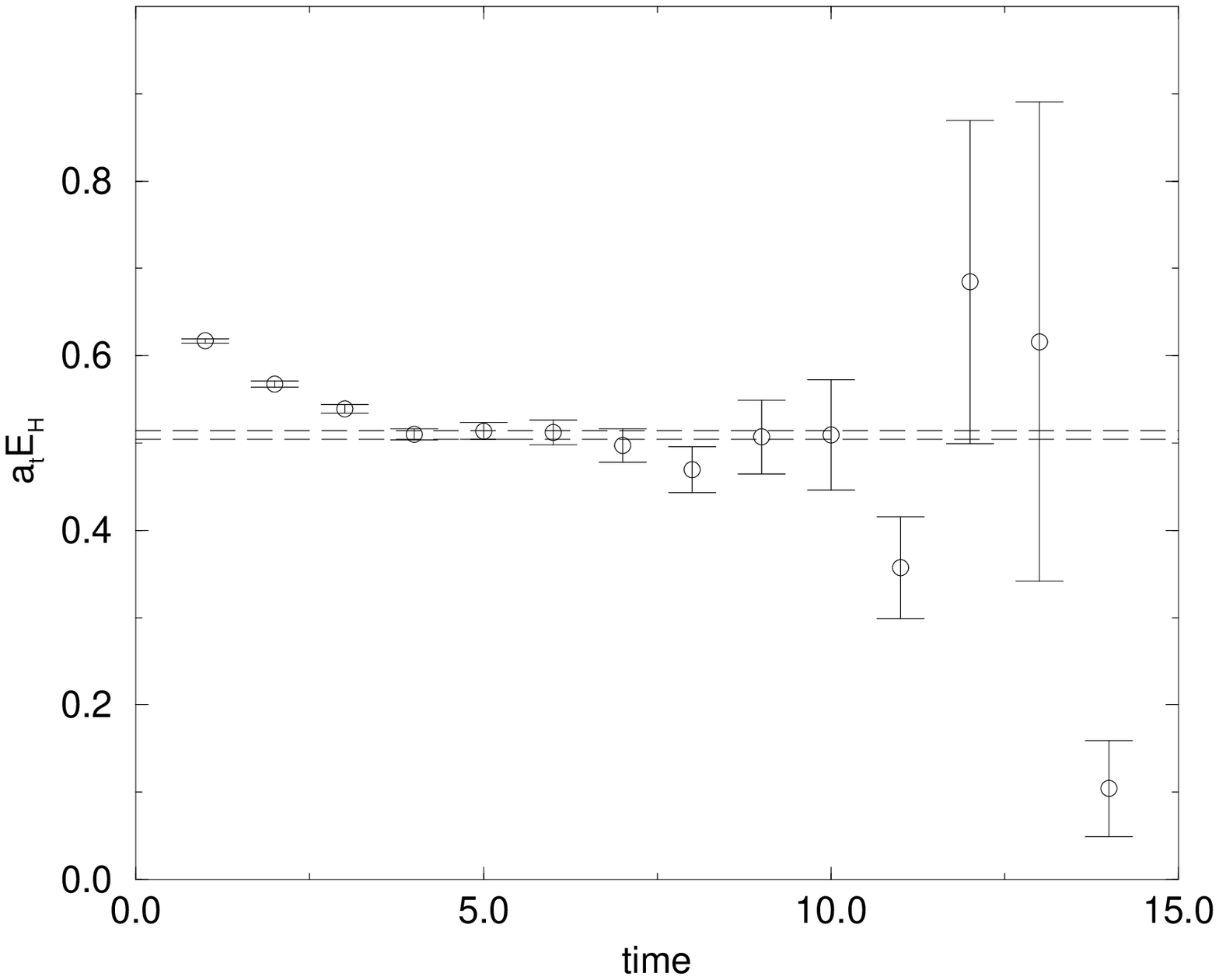,height=80mm}
\enc
\caption{\label{effm}\small
Effective mass plot for the $1^{-+}$ hybrid with $c_1=-1$. The dashed lines
show the upper and lower bounds of the fit. The plateau extends from t=4--10.
}
\enf

\befh
\bec
\epsfig{file=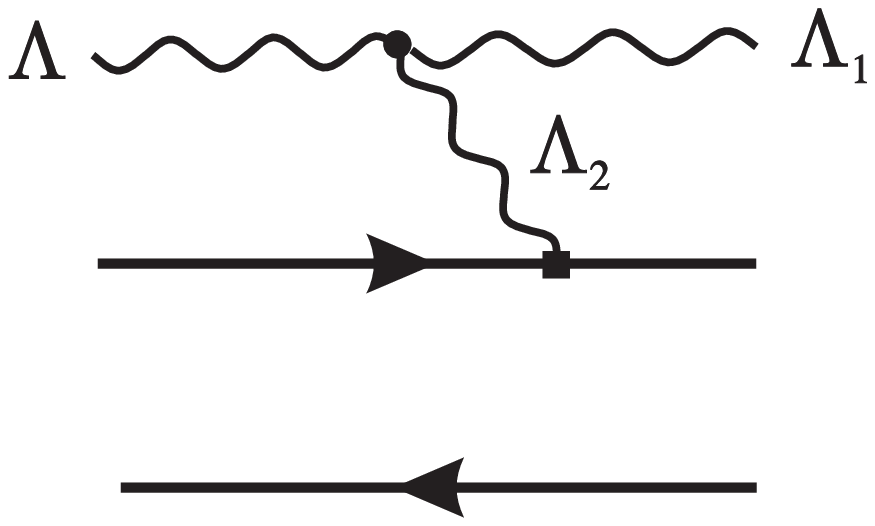,height=40mm}
\enc
\caption{\label{tgluon}\small
A representative graph for the second-order perturbation theory calculation discussed 
in section \ref{BOP} for the mass shift due to the inclusion of the $\bsig.\bB$ term 
in the NRQCD Hamiltonian (\ref{h_nrqcd}). This spin operator is represented by the filled box
and terms in the spin independent part of the interaction Hamiltonian are represented by
filled circles. This contribution is to the linear term in $c_1$ in eq(\ref{fit}) and is 
$O(\a_S)$. 
}
\enf

\befh
\bec
\epsfig{file=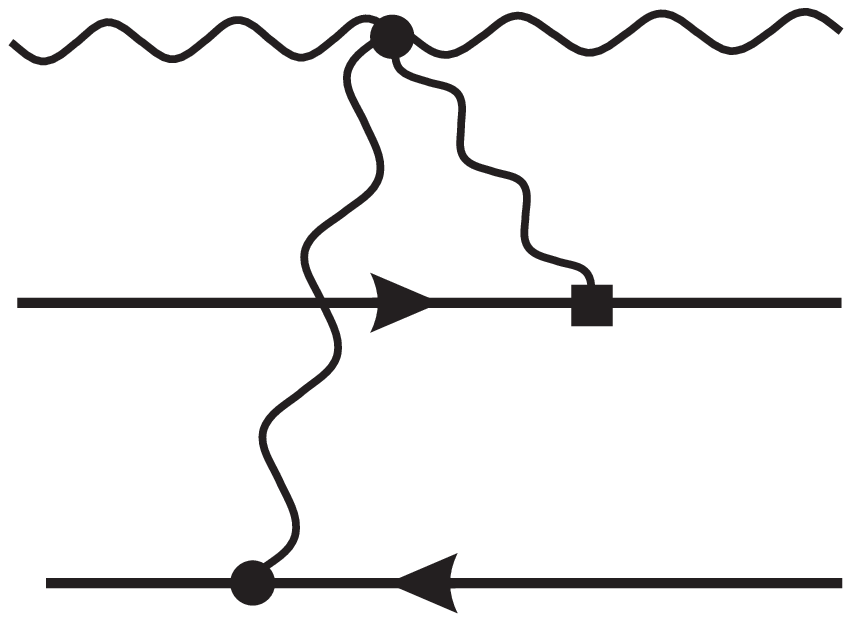,height=40mm}
\enc
\caption{\label{fgluon}\small
A higher-order, $O(\a_S^2)$, perturbation contribution to the linear term in $c_1$ 
in eq(\ref{fit}). Symbolism is the same as in Fig. \ref{tgluon}.
}
\enf

\befh
\bec
\epsfig{file=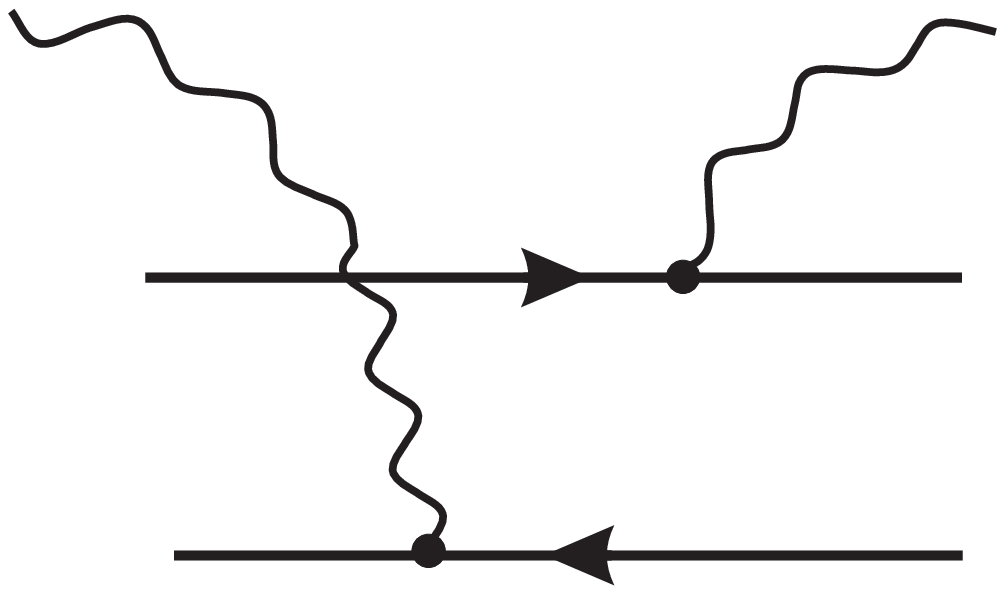,height=40mm}
\enc
\caption{\label{mixing}\small
A representative graph for the mixing contribution to the hybrid-hybrid correlator from
quarkonium intermediate states. 
}
\enf

\end{document}